\documentclass[prb,twocolumn,aps,showpacs,fixfloats]{revtex4}
\usepackage{graphicx}
\usepackage{bm}
\usepackage{amsmath,amssymb}
\usepackage{subfigure}
\usepackage{float}
\usepackage{latexsym}
\usepackage{color}
\usepackage{enumerate}
\usepackage{pdfpages}
\usepackage{tikz}
\usepackage{hyperref}
\usepackage{graphicx}
\usepackage{bm}
\usepackage{amssymb} 
\usepackage{amsmath}
\usepackage{subfigure}
\usepackage{relsize}

\begin{document}
\newcommand{\s}{\scriptscriptstyle}
\newcommand{\uu}{\uparrow \uparrow}
\newcommand{\ud}{\uparrow \downarrow}
\newcommand{\du}{\downarrow \uparrow}
\newcommand{\dd}{\downarrow \downarrow}
\newcommand{\ket}[1] { \left|{#1}\right> }
\newcommand{\bra}[1] { \left<{#1}\right| }
\newcommand{\bracket}[2] {\left< \left. {#1} \right| {#2} \right>}
\newcommand{\vc}[1] {\ensuremath {\bm {#1}}}
\newcommand{\tr}{\text{Tr}}
\newcommand{\Trans}{\ensuremath \Upsilon}
\newcommand{\Refl}{\ensuremath \mathcal{R}}


\title{Interaction-induced long-time tail of a nonlinear ac absorption in a localized  system:
 a relay-race mechanism}

\author{Rajesh K. Malla   and M. E. Raikh}

\affiliation{ Department of Physics and
Astronomy, University of Utah, Salt Lake City, UT 84112}

\begin{abstract}

In conventional solid-state electron systems with localized states the ac absorption is linear since the inelastic widths of the energy levels exceeds the drive amplitude. The situation is different in the systems of cold atoms in which phonons are absent. Then even a weak drive leads to saturation of the ac absorption within resonant pairs, so that the population of levels oscillates with the Rabi frequency. We demonstrate that, in the presence of weak dipole-dipole interactions,
the response of the system acquires a long-time component which oscillates with frequency much smaller than the Rabi frequency. The underlying mechanism of this long-time behavior is that the fields created in the course of the Rabi oscillations serve as resonant drive for the {\em second-generation} Rabi oscillations
in pairs with level spacings close to the Rabi frequency. The frequency of the second-generation oscillations is of the order of interaction strength. As these oscillations develop, they can initiate the next-generation Rabi oscillations, and so on.
Formation of the second-generation oscillations 
is  facilitated by the non-diagonal component of the    dipole-dipole interaction tensor.
\end{abstract}

\maketitle

\section{Introduction}
A  transparent physical picture of absorption of the ac electric field in a system with localized electron states was proposed by N. F. Mott.\cite{Mott}
According to Mott, absorption takes place within 
pairs of states with energy spacing $\hbar\omega$, where $\omega$ is the driving frequency.
Frequency dependence of the ac conductivity within this picture is $\sigma(\omega)\propto  \omega^2\ln^2\omega$ (in one dimension), 
where one power of $\omega$ comes from the photon energy, while the other comes from the  
restriction that the pair is singly occupied. Finally, $\ln\omega$
comes from the overlap integral between initial and final states. Later, Mott's formula was rigorously derived from the Kubo linear-response formalism by V.~ L. Berezinskii\cite{Berezinskii}.

The condition of applicability of the linear response is that inelastic widths of the localized levels are much bigger than the absorption matrix element. This condition is satisfied in conventional solid-state systems where inelastic widths are due to the phonon emission.

Regime of strong ac drive, opposite to the linear response, can be realized in cold-atom systems, where phonons are absent. The ac drive in these systems is implemented by the synchronous modulation of the intensity of laser beams which create a quasi-random  1D on-site energy profile.\cite{NaturePhysics}

Possibility to realize the regime of  strong drive in a localized system without thermal bath raises a number of conceptual questions which, with rare exceptions,\cite{GefenThouless,Kravtsov,Kottos} were not addressed in earlier studies. These questions can be conventionally  divided into three groups:

({\em i}) On the single-particle level,
\cite{SondhiNaturePhysics,French,Bhatt,AA,Chalker} the fundamental question is: does the localization persist in the presence of strong drive, 
when electron states evolve into the Floquet eigenstates? Anderson localization is the result of interference of the backscattering amplitudes in the course of multiple scattering\cite{Berezinskii}. Floquet states can be viewed as combination of satellites with energies separated by $n\hbar\omega$.
Development of satellites upon increasing drive leads to the new channels of interference, and thus suppresses the localization, like
in multichannel wires.   

({\em ii}) Another physical mechanism relevant for nonlinear ac response of localized 
non-interacting systems 
is the adiabatic Landau-Zener transitions.\cite{Bhatt,Chalker}  This mechanism comes into play  when the drive is strong and slow. In this limit, the effect of drive can be viewed as 
periodic modulation of energies of the 
localized states.\cite{Wilkinson1988} 
As the levels corresponding to neighboring states slowly pass by each other, an electron can adiabatically change the level.
This, in turn, can  lead to the long-time 
component of the ac absorption.\cite{Demler1}
Spreading of electron due to the level crossings
illustrates the tendency of drive to suppress the localization.

%
%



({\em iii}) The third group of papers is the most numerous, see e.g. Refs.\onlinecite{Demler1,Huse1,AbaninDriven,AbaninDriven1,Moessner,Bukov,Moessner1,Lindner},                        and addresses the physics of ac driven
localized interacting systems. They are focused
on the dynamics of heating and on the long-times  properties of non-equilibrium state. In particular, the question of interest is whether or not the long-time behavior of interacting 
many-body system is sensitive to its initial state.

When the drive amplitude is much smaller than
the drive frequency, $\omega$, resonant pairs get
saturated after the time of the order of the inverse Rabi period. Higher harmonics in the pair
dynamics are small in this regime.\cite{Shirley}
Landau-Zener transitions also do not take place when the drive is fast. It is argued in Ref. \onlinecite{Demler1} that  long-time dynamics 
in this limit
is due to interaction between the pairs. Namely,
a group of $n$ interacting pairs can be engaged
into collective Rabi oscillations, whose frequency is proportional to $n$-th power of drive.

In the present paper we propose an alternative 
mechanism of long-time dynamics in a system of  weakly-interacting pairs under a weak drive.
 Namely, a saturated pair, executing the Rabi oscillations, creates a field which plays the role of drive for a distant pair, thus causing the {\em second-generation} Rabi oscillations.
At resonance, the level  spacing of the second-generation pair is equal to the Rabi frequency of
the first-generation pair. If this frequency is
much smaller than the pair-pair interaction,
then the second-generation Rabi oscillations are
slow. We perform statistical averaging analytically and find the slow component of the
absorption.  
  
\begin{widetext}

\begin{figure}[h!]
\includegraphics[scale=0.25]{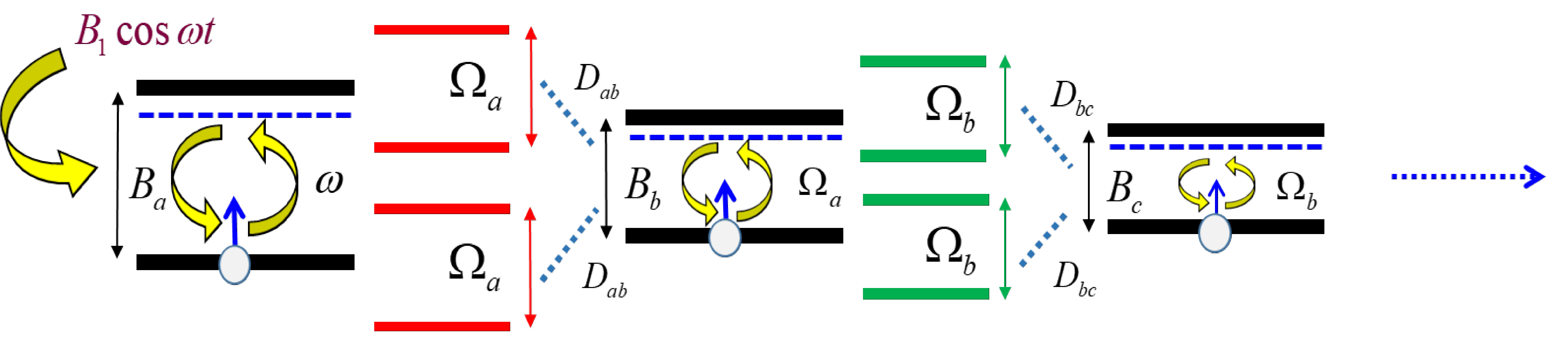}
\caption{(Color online) Schematic illustration of the relay-race mechanism: resonant drive with frequency, $\omega$, engages spin $a$ into the Rabi oscillations. As a result, the 
Zeeman levels of $a$ get split by $\Omega_a$, the Rabi frequency. If $\Omega_a$ is close to the Zeeman splitting of spin $b$, then the Rabi oscillations of $a$ will serve as a resonant drive for $b$ via the non-diagonal
component of the dipole-dipole interaction,
$D_{ab}$.
If, in turn, the Rabi frequency, $\Omega_b$, of
the second-generation Rabi oscillations  
is close to the Zeeman splitting of spin $c$, the third-generation Rabi oscillations are
initiated as a result of dipole-dipole interaction, $D_{bc}$, and so on.     }
\label{frabi}
\end{figure}

\end{widetext}
\section{Dynamics of two interacting driven spins}

To illustrate the proposed mechanism we employ the simplest model. Namely,
as is common in the literature, we employ the language of spins to describe two-level systems and, correspondingly, the ac magnetic field to describe the drive. Consider two spins, $a$ and $b$, subject to magnetic fields $B_a$ and $B_b$,
respectively. Since the drive amplitude, $B_1$,
is much smaller than $\omega$, the rotating-wave
approximation applies. Then the Hamiltonian of the pair reads

\begin{multline}
\label{Hamiltonian}
\hat{H}=\frac{B_a}{2} S_z^a+\frac{B_b}{2} S_z^b+2B_1\left(S_x^a +S_x^b \right)\cos \omega t
\\
+2B_1\left(S_y^a +S_y^b \right)\sin \omega t
-D\frac{3({\bf S}^a\cdot{\bf r})({\bf S}^b\cdot{\bf r})-({\bf S}^a\cdot{\bf S}^b)r^2}{r^5},
\end{multline}
where we have incorporated the dipole-dipole interaction with magnitude, $D$, while ${\bf r}$
is the vector-distance between $a$ and $b$.

In the absence of interaction, only spin 
$a$ is in resonance with the drive 
$|\omega-B_a|\ll \omega$, while spin $b$ is
off resonance, $B_b\ll \omega$, and does not
respond to the drive. Components $S_a^x$ and $S_a^y$ of the driven spin $a$ oscillate with
frequencies close to $\omega$. Thus, the fields
produced by these components on spin $b$ via dipole-dipole interaction, do not induce the dynamics of $b$. 
On the other hand, the $z$-component of spin $a$
oscillates with much smaller frequency
\begin{equation}
\label{RabiFrequency}
\Omega_a=\left[B_1^2+(\omega-B_a)^2 \right]^{1/2}.
\end{equation}
These slow oscillations of $S_a^z$ 
translate into the field acting on $b$.
The field generated by the 
$z$-$z$ component
of the interaction is also inefficient,
since spin $b$ is already directed along $z$.  
Spin $b$ can be set in motion via the 
$z$-$x$
and $z$-$y$
components of the dipole-dipole interaction when $\Omega_a$ is close to $B_b$. This is why we will keep only the 
$z$-$x$ component.

Summarizing, the relevant components of the
field acting on $a$ are $(B_1 \cos \omega t, B_1 \sin \omega t, B_a +D S_x^b)$, while the field
acting on $b$ has an $x$-component,  equal 
to $D S_z^a$, and $z$-component, $B_b$.

The equations of motion for the projections 
of $a$ which follow from $\frac{d\bf{S}}{dt}= \bf{B}\times \bf{S}$, read


%
%

\begin{eqnarray}
\label{Spina}
\frac{dS_x^a}{dt}&=&B_1 \sin \omega t S_z^a-\left(B_a+D S_x^b \right)S_y^a,\\
\frac{dS_y^a}{dt}&=&-B_1 \cos \omega t S_z^a+\left(B_a+D S_x^b \right)S_x^a,\\
\frac{dS_z^a}{dt}&=&B_1 \cos \omega t S_y^a -B_1 \sin \omega t S_x^a,
\end{eqnarray}
while the equations of motion for the components of spin $b$ have the form
\begin{eqnarray}
\label{Spinb}
\frac{dS_x^b}{dt}&=&-B_b S_y^b,~~
\frac{dS_y^b}{dt}=-DS_z^aS_z^b+B_bS_x^b,\\
\frac{dS_z^b}{dt}&=&DS_z^a S_y^b.
\end{eqnarray}

To analyze the coupled equations of motion for 
$a$ and $b$ it is convenient to cast them into the integral form.
First, we express $S_x^a$ and $S_y^a$ in terms
of $S_z^a$ and $S_x^b$. Substituting the result
into the equation for $S_z^a$ and taking into account the initial condition $S_z^a(0)=1$, we get
\begin{equation}
\label{reduced1}
\frac{d S_z^a}{dt}
=\hspace{-4mm}-B_1^2\int\limits_0^t dt' S_z^a (t')
\times\cos\Big[(\omega -B_a)(t-t') - D\int\limits_{t'}^{t} dt''S_x^b(t'')\Big].
\end{equation}
Similarly, we express $S_x^b$ and $S_y^b$ in 
terms of $S_z^a$ and $S_z^b$ and, using $S_z^b(0)=1$, substitute them into the equation for
$S_z^b$. This yields
\begin{equation}
\label{reduced3}
\frac{dS_z^b}{dt}=-D^2\int\limits_0^t dt' S_z^b(t')\left[S_z^a(t)S_z^a(t')\right]\cos B_b(t-t').
\end{equation}

To get the closed system, we also invoke
the expression for $S_x^b$ obtained in the course of solving the system (\ref{Spinb}). 
 
\begin{equation}
\label{reduced2}
S_x^b(t)=-\frac{D}{B_b}(1-\cos B_b t)+D \int\limits_0^t dt' S_z^a(t')S_z^b(t')\sin B_b(t-t').
\end{equation}
Three equations (\ref{reduced1}), (\ref{reduced3}), and (\ref{reduced2})
describe fully the dynamics of both spins.

For $D=0$, spin $b$ points along $z$, while spin $a$ executes the
Rabi nutations. In course of these nutations $S_z^a(t)$ follows
the seminal Rabi formula
\begin{equation}
\label{Sza}
S_z^a(t)=\frac{\left(\omega - B_a\right)^2}{\Omega_a^2}+ \frac{B_1^2}{\Omega_a^2}\cos\Omega_a t,
\end{equation}
which also follows from equation (\ref{reduced1}).

For a finite $D$ spin $b$ is also set into motion. This motion causes a ``feedback" on spin $a$, reflected by the term proportional to $D$ in the argument of cosine.
Most importantly, comparison of (\ref{reduced1}) and (\ref{reduced3}) quantifies our main message that the motion of spin $a$ plays the role of drive for the spin $b$; the role of the driving field is played by $D S_z^a(t)$.

\begin{figure}
\includegraphics[scale=0.35]{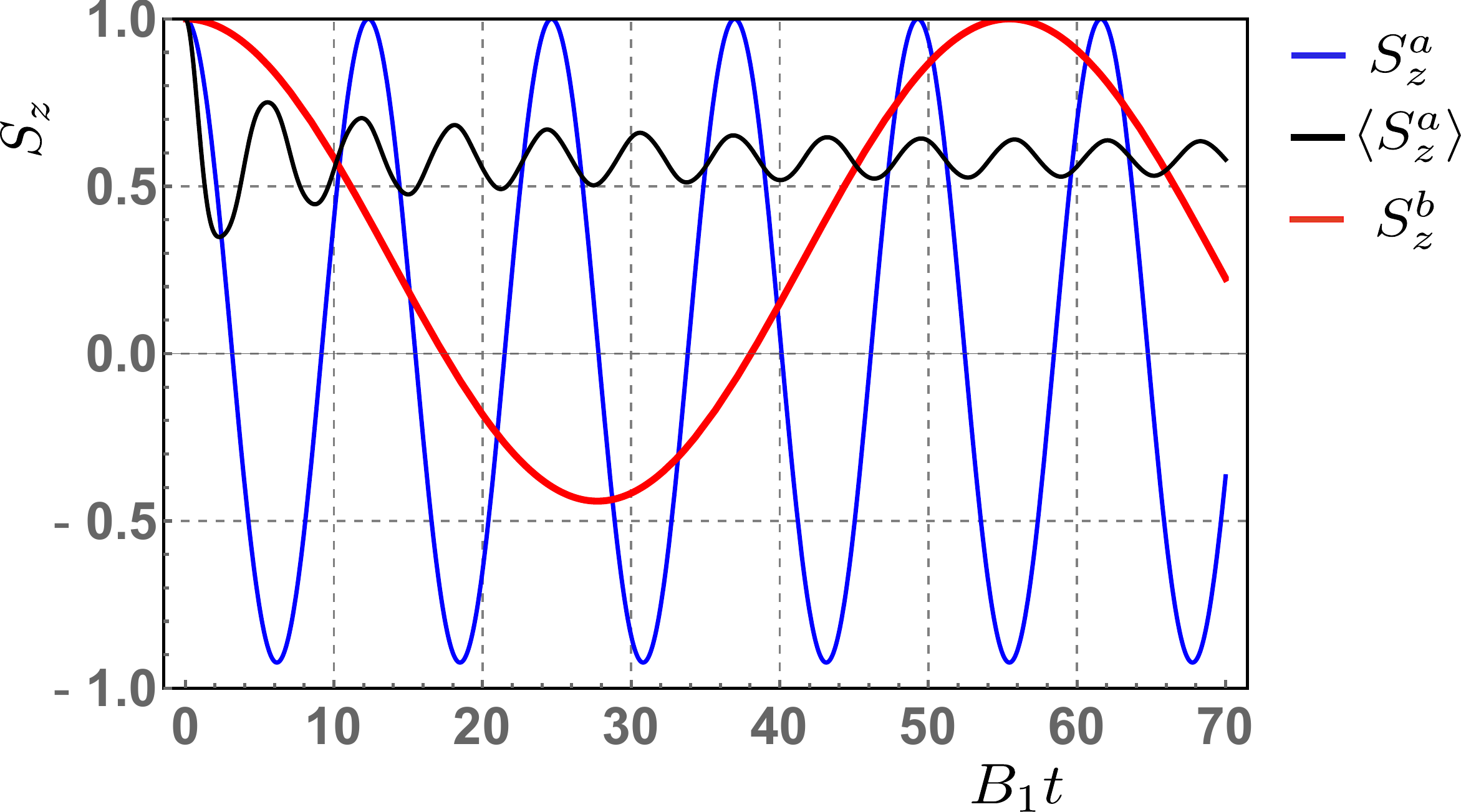}
\caption{(Color online) Numerical example illustrating the formation of the long-time component of the ac response. First generation
of the Rabi oscillations (blue) is plotted from (\ref{Sza}) for the drive frequency $\omega=2B_1$ and the Zeeman splitting 
$B_a=1.8B_1$. Assuming that the Zeeman energies are homogeneously distributed between
$0$ and $1.5\omega$, the ensemble-averaged
$\langle S_z(t)\rangle$ is calculated from (\ref{disorder}) and is plotted with black.
Second-generation Rabi oscillations are 
shown with red. They are calculated from (\ref{Szb}), (\ref{second}) in which we chose the Zeeman
energy of spin $b$ to be  
$B_b=0.5B_1$ and the magnitude of interaction to be   $D_0=0.05B_1$.   }
\label{f2}
\end{figure}

For spin $a$ the resonant drive corresponds to the frequency $\omega =B_a$. Under this condition, $S_z^a$ oscillates with frequency $B_1$.
Thus, for spin $b$,  the resonant condition  is $B_1=B_b$. We also expect
that, within a factor, the nutation frequency of spin $b$ at resonance is equal to $D$, as illustrated in
Fig. \ref{f2}. Note that the nutation frequency, $D$, of the second-generation Rabi oscillations does not depend on $B_1$. However, this is valid only at exact resonance.
We will see below that, for any small deviation from resonance, both the amplitude and the
frequency of the second-generation Rabi oscillations acquire the $B_1$-dependence. In particular, the
amplitude vanishes in the limit $B_1\rightarrow 0$.

Upon substituting (\ref{Sza}) into (\ref{reduced3}), the product $S_z^a(t)S_z^a(t')$ assumes the form
\begin{multline}
\label{tt'}
 S_z^a(t)S_z^a(t')= \frac{\left(\omega - B_a\right)^4}{\Omega_a^4}+ \frac{B_1^4}{\Omega_a^4}\cos\Omega_a t \cos\Omega_a t' \\
+ \frac{\left(\omega - B_a\right)^2B_1^2}{\Omega_a^4} \left(\cos \Omega_a t +\cos \Omega_a t' \right)
\end{multline}
The second term of (\ref{tt'}) 
generates the sum 
$\cos \Omega_a (t+t')+\cos \Omega_a(t-t')$. 
It is the second cosine that acts as a 
resonant drive for the spin $b$.
Keeping only $\cos \Omega_a(t-t')$-term in (\ref{reduced3}), we get
\begin{equation}
\label{Szbmodified}
\frac{dS_z^b}{dt}=-\frac{D^2 B_1^4}{4 \Omega_a^4}\int\limits_0^t dt' S_z^b(t')\cos\left[ \left(\Omega_a-B_b\right)(t-t')\right].
\end{equation}
This equation has a solution
\begin{equation}
\label{Szb}
S_z^b= \frac{(\Omega_a-B_b)^2}{\Omega_b^2}+\frac{B_2^2}{\Omega_b^2} \cos \Omega_b t,
\end{equation}
where the ``second-generation" drive and the second-generation
Rabi frequency are defined as
\begin{equation}
\label{second}
 B_2=\frac{D B_1^2}{2 \Omega_a^2},~~~~~~
\Omega_b=\left[B_2^2 + (\Omega_a-B_b)^2 \right]^{1/2}.
\end{equation}
Equations (\ref{Szb}) and (\ref{second})  constitute
the main result of the present paper. We analyze this result below.

\section{Statistical averaging, long-time tail of the ac absorption}
The energy absorbed is proportional to averaged $S_z^b(t)$. We should average $S_z^b(t) $ over $B_a$, $B_b$ and over $D$, which is equivalent to averaging over distances to the neighbors. We notice that $B_a$ enters into the $S^b_z(t)$ only in combination $(B_a-\omega)^2+B_1^2$, so that $B_a$ does not affect the result of averaging.

We now average $S_z^b(t)$, from equation (\ref{Szb}), assuming the  density of spin states, $g$, to be constant. 
The averaging amounts to the two-fold integral 
\begin{multline}
\label{average1}
1-\langle S_z^b(t)\rangle =g\hspace{-0.5mm}\int\limits_a^\infty dr 4\pi r^2\hspace{-1mm}\int\limits_0^\infty \hspace{-1mm} dB_b \frac{\left(\frac{D(r)B_1^2}{4\Omega_a^2}\right)^2}{\left(\frac{D(r)B_1^2}{4\Omega_a^2}\right)^2+(\Omega_a-B_b)^2}
\\
\times 2\sin^2\Bigg\{\Bigg[\left(\frac{D(r)B_1^2}{4\Omega_a^2}\right)^2+\left(\Omega_a-B_b\right)^2\Bigg]^{1/2}\hspace{-4mm} \hspace{3mm}\frac{t}{2}\hspace{1mm}\Bigg\},
\end{multline}
where $a$ is the distance to the neighboring spin.
Angular averaging is not important, so we choose 
\begin{equation}
D(r)=D_0\left(\frac{a}{r}\right)^3,
\end{equation}
where $D_0$ is the dipole-dipole interaction between two neighbors. 

To proceed further, it is convenient to introduce, instead of variables $r$ and $B_b$, new variables $u$ and $v$ defined as
\begin{equation}
\label{substitution}
r = \left(\frac{D_0 B_1^2 a^3}{8\Omega_a^2 u}t\right)^{1/3}  ,~~~B_b=\Omega_a-\frac{2v}{t}.
\end{equation}
Then the integral assumes the form
\begin{equation}
\label{3D}
1-\langle S_z^b(t)\rangle =\frac{2\pi ga^3 D_0B_1^2}{3\Omega_a^2} \int\limits_{0}^{D_0t/2} \hspace{-1.1mm} du
\int\limits_{-\Omega_a t}^\infty \hspace{-1.1mm} dv~~ \frac{\sin^2\left(u^2+v^2\right)^{1/2}}{u^2+v^2}.
\end{equation}
In the long-time limit $D_0t \gg 1$, $\Omega_a t \gg D_0t$, we can replace $\Omega_a t$ with infinity in the lower limit of the $v$-integral.  In order to see the asymptotic behavior of $\langle S_z^b(t)\rangle$, we substitute $v$ as, $v=\frac{u}{\tan \psi}$. With new variable, $\psi$, the integral (\ref{3D}) can be rewritten as
\begin{equation}
\label{3D1}
1-\langle S_z^b(t)\rangle\Bigg|_{3D} =\frac{4\pi ga^3 D_0B_1^2}{3\Omega_a^2} \int\limits_{0}^{D_0t/2} \frac{du}{u} \Phi(u),
\end{equation}  
with $\Phi(u)$ defined as 
\begin{equation}
\label{phiu}
  \Phi(u)  =\int\limits_{0}^{\pi/2} d\psi~~ \sin^2\left(\frac{u}{\sin\psi}\right).
\end{equation}
In one and two dimensions, the corresponding expressions for $1-\langle S_z^b(t)\rangle$ are similar to (\ref{3D1}) and read
\begin{eqnarray}
1-\langle S_z^b(t)\rangle\Bigg|_{1D}\hspace{-2mm}=\frac{4 gaD_0}{3\left(\frac{D_0t}{2}\right)^{2/3}}
\left(\frac{B_1^2}{4\Omega_a^2}\right)^{1/3}\hspace{-1mm} \int\limits_{0}^{D_0t/2} \frac{du}{u^{1/3}} \Phi(u),\hspace{2mm}\label{1}\\
1-\langle S_z^b(t)\rangle\Bigg|_{2D}\hspace{-4mm}=\frac{8\pi ga^2D_0}{3\left(\frac{D_0t}{2}\right)^{1/3}}
\left(\frac{B_1^2}{4\Omega_a^2}\right)^{2/3}\hspace{-1mm} \int\limits_{0}^{D_0t/2} \hspace{-2mm}\frac{du}{u^{2/3}} \Phi(u).\hspace{3mm}\label{2D}
\end{eqnarray}

The function $\Phi(u)$ in (\ref{phiu}) can be calculated analytically in two limits.  For small $u\ll 1$, only
small $\psi\sim u$ contribute to the integral. This allows
to replace $\sin\psi$ by the argument and extend the integration to infinity. The resulting integral can be calculated explicitly, and one gets
\begin{equation}
\label{small}
\Phi(u)\Big|_{u\ll 1} =\frac{\pi u}{2}.
\end{equation}
For large $u\gg 1$, the typical
argument of $\sin^2$ is big, so that $\sin^2$ can be replaced
by $1/2$, leading to 
$\Phi(u)\big|_{u\gg 1}\approx \frac{\pi}{4}$.
The leading $u$-dependent correction comes 
from the vicinity of $\psi=\frac{\pi}{2}$, and
thus oscillates with $u$. The asymptote has the
form
 
\begin{equation}
\label{large}
\Phi(u)\Big|_{u\gg 1}=\frac{\pi}{4}-\left(\frac{\pi}{8}\right)^{1/2}\frac{\cos\left(2u+\frac{\pi}{4}\right)}{u^{1/2}}.
\end{equation}

Using (\ref{small}) and (\ref{large}), we find the behavior  of  $\langle S_z^b(t)\rangle$ in three dimensions 
\begin{multline}
\label{Result1}
1-\langle S_z^b(t)\rangle =\frac{8\pi ga^3 D_0B_1^2}{\Omega_a^2}\times
\left\{\begin{array}{ll}
D_0t,~~ &\frac{D_0t}{2}\ll 1, \\
\frac{1}{2}\ln (D_0t),~~& D_0t \gg1 
.
\end{array} \right. 
\end{multline} 
Note that, at $D_0t \ll 1$, the average 
$\langle S_z^b(t)\rangle$ decreases {\em linearly} with
time. This is despite the fact that, for any {\em given}
spin $b$, the time deviation of $S_z$ from $S_z=1$  is quadratic.
The reason is that, if one expands (\ref{average1}) at small $t$, then the integral over $B_b$ will diverge.
Overall, the characteristic time-scale for the change
of  $\langle S_z^b(t)\rangle$ is $D_0^{-1}$. 

In derivation of the expression for  $S_z^b(t)$, we have already assumed
that $D_0\ll B_1$ when we neglected the feedback of $b$ on $a$. Now we see that the same assumption insures that the
the evolution of the ensemble-averaged $S_z^b(t)$ is slow.  
 
\begin{figure}
\includegraphics[scale=0.3]{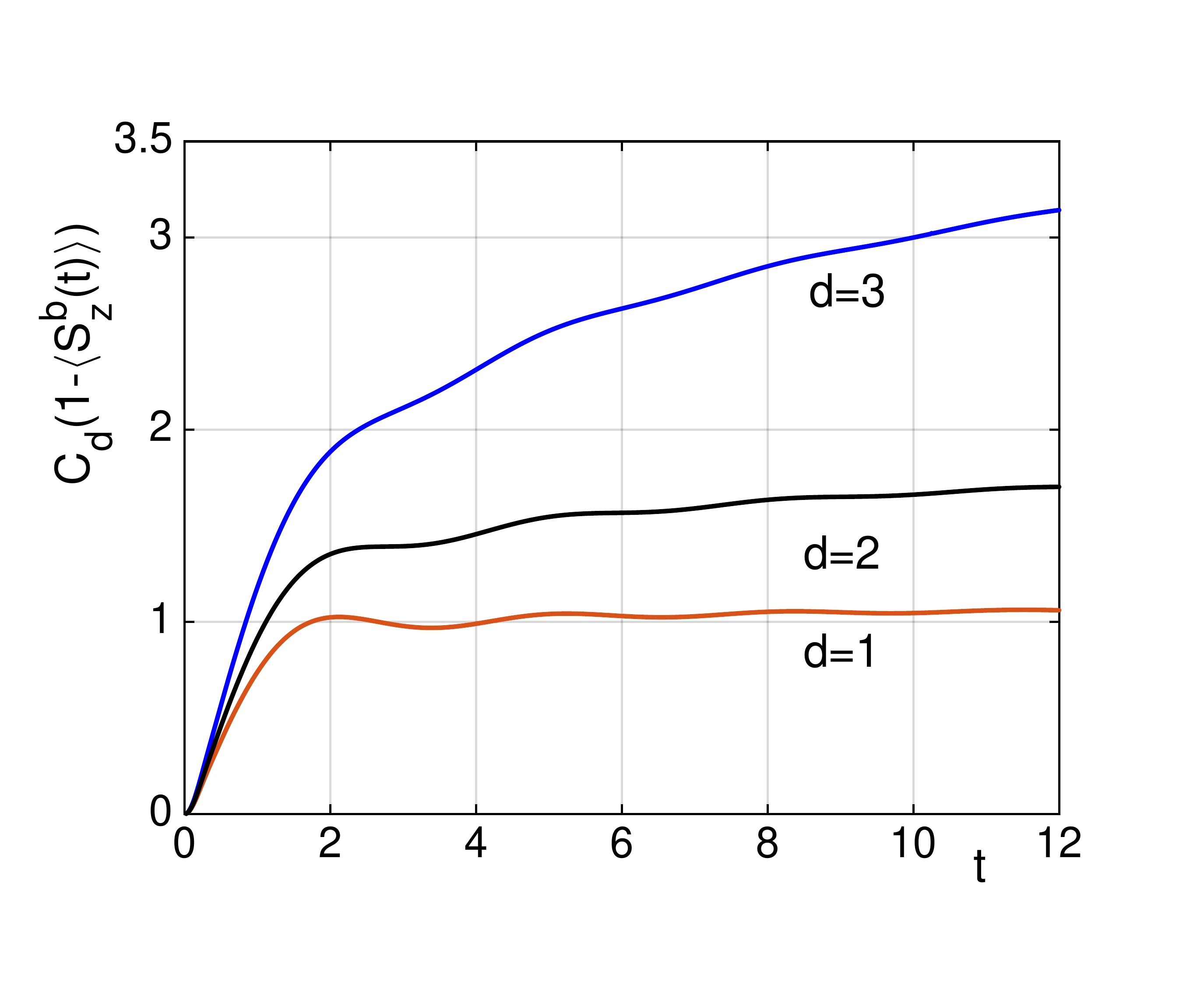}
\caption{(Color online)  The average $1-\langle S_z^b(t)\rangle$ is  plotted 
versus dimensionless time, $D_0t$, from equations (\ref{1}), (\ref{2D}), and (\ref{3D1}) corresponding to one, (red) two, (black) and three (blue) dimensions. The values of the coefficient $C_d$ are:  $\frac{4 g aD_0}{3}
\left(\frac{B_1^2}{4\Omega_a^2}\right)^{1/3}$, $\frac{8\pi g a^2D_0}{3}
\left(\frac{B_1^2}{4\Omega_a^2}\right)^{2/3}$, and $\frac{4\pi g a^3 D_0B_1^2}{3\Omega_a^2}$ for one, two and three dimensions, respectively. }
\label{f3}
\end{figure}

In Fig. \ref{f3}, we show $\langle S_z^b(t)\rangle$ calculated numerically from the equation (\ref{3D1}). Logarithmic behavior is evident. One can also distinguish 
weak oscillations on the background of log profile. These oscillations become more pronounced in lower dimensions.
For example, in one dimension, the equation (\ref{1}) can be written in the following form:
\begin{equation}
\label{11}
1-\langle S_z^b(t)\rangle = \frac{4 gaD_0}{3\left(\frac{D_0t}{2}\right)^{2/3}}
\left(\frac{B_1^2}{4\Omega_a^2}\right)^{1/3}F(D_0t),
\end{equation}
where
\begin{equation}
\label{F}
F(D_0t)=\int\limits_{0}^{D_0t/2} \frac{du}{u^{1/3}} \Phi(u).
\end{equation}
If we substitute the leading asymptote, $\Phi(u)=\frac{\pi}{4}$, we will find that $\langle S_z^b(t) \rangle$
is time-independent. In order to capture the time dependence, we add and subtract $\pi/4$ from $\Phi(u)$. Then (\ref{F}) takes the form
\begin{equation}
\label{F1}
F(D_0t)=\frac{3\pi}{8}\left(\frac{D_0t}{2}\right)^{2/3 }+\int\limits_{0}^{D_0t/2} \frac{du}{u^{1/3}} \left(\Phi(u)-\frac{\pi}{4}\right).
\end{equation}
At long times, the first term describes the leading contribution, while the second term
saturates. This saturation is accompanied by
the oscillations. To establish the form of this oscillating
correction, it is convenient to present the
integral $\int\limits_{0}^{D_0t/2}du$ as the
difference of integrals $\int\limits_{0}^{\infty}du$ and $\int\limits_{D_0t/2}^{\infty}du$.
Then, in the integral, $\int\limits_{D_0t/2}^{\infty}du$, we can use the
oscillating term from the large-$u$ asymptote in (\ref{large}). This generates the following correction to $F(D_0t)$:
$$
-\left(\frac{\pi}{8}\right)^{1/2}\frac{\sin\left(D_0t+\frac{\pi}{4}\right)}{2\left(\frac{D_0t}{2}\right)^{5/6}}.
$$
Substituting this correction into (\ref{11}) indicates that the amplitude of oscillations in $\langle S_z^b(t)\rangle$ falls off as $t^{-3/2}$. Numerical plots for $d=1$ and $d=2$ in Fig. \ref{f3} confirm the saturation of $\langle S_z^b(t)\rangle$ at long times, which is accompanied by slow-decaying oscillations. As follows from analytics and numerics, the small time behavior of $1-\langle S_z^b(t)\rangle$ is linear in one and two dimensions as well.

\section{The role of feedback}

Unconventionally, we find that the coupling of two spins via dipole-dipole interaction
is {\em ``unidirectional"}: spin $a$ drives spin $b$, while the feedback effect of $b$ on $a$
is negligible under the condition $D\ll B_1$. On the other hand, it is the domain $D \ll B_1$,
which is of interest, since it is in this domain where the long-time tail of the ac absorption develops. If we take into account that spin $b$ is dipole-dipole coupled to spin ``$c$",  see Fig. \ref{frabi}, then $b$ will drive $c$ under the resonant condition, but with negligible  feedback. This is why we identify this spin dynamics with relay-race.

To estimate the effect of feedback, we take
the expression  for  $S_z^a(t)S_z^b(t)$ obtained in the lowest order and substitute it into (\ref{reduced2}). 
The expression for $S_z^a(t)S_z^b(t)$ has the form similar to the expression for $S_z^a(t)S_z^a(t')$ in (\ref{tt'})  with
$t'=t$
\begin{widetext}
\begin{equation}
\label{wide}
S_z^a(t)S_z^b(t)= \frac{\left(\omega - B_a\right)^2\left(\Omega_a - B_b\right)^2}{\Omega_a^2\Omega_b^2}+ \frac{B_1^2 B_2^2}{\Omega_a^2\Omega_b^2}\cos\Omega_a t \cos\Omega_b t 
+ \frac{\left(\Omega_a - B_b\right)^2B_1^2}{\Omega_a^2\Omega_b^2}\cos \Omega_a t +\frac{\left(\omega - B_a\right)^2B_2^2}{\Omega_a^2\Omega_b^2}\cos \Omega_b t. 
\end{equation}
\end{widetext}
As seen from (\ref{reduced2}), $S_x^b$ contains a ``free precession"
term,  $\frac{D}{B_b}(1-\cos B_bt)$, and the ``drive-induced" term.
The role of the free precession term is the shift of resonance
$\omega=B_a$. Indeed, substituting this term into the argument of
cosine in (\ref{reduced1}), and assuming that $B_bt$ is big (or, equivalently, that $\Omega_at$ is big), 
results in replacement of $(\omega-B_b)$ by $(\omega-B_b-\frac{D^2}{B_b})$, i.e. the corrected resonance condition is 
\begin{equation}
\label{corrected}
\omega=B_b\left(1+\frac{D^2}{B_b^2}\right). 
\end{equation}
Since the relevant
value of $B_b$ is the Rabi frequency, $B_1$, we conclude that
the shift is relatively small under the condition  $B_1\gg D$, 
which coincides with the condition that the second-generation
Rabi oscillations are slow. 

We neglected $\cos B_bt$ in free precession term because it leads
to the oscillating contribution, $\frac{D^2}{B_b^2}\left(\sin B_bt 
-\sin B_bt'\right)$ in the argument of cosine in (\ref{reduced1}).
This oscillating contribution results to effective renormalization
of the drive amplitude\cite{Glenn1} 
$B_1\rightarrow B_1J_0\left(\frac{D^2}{B_b^2}\right)$, where $J_0$
is the Bessel function. This renormalization is small by virtue of
the same condition, $D\ll B_b\sim B_1$.

We now turn to the effect of feedback from  the drive-induced term.
As we have established in the course of statistical averaging, the
second-generation Rabi oscillations essentially saturate at times $\sim  \frac{1}{D} $.
On the other hand, we do not expect significant feedback at times smaller that the
period of the first-generation Rabi oscillations. This simplifies our task by restricting the time 
to the interval
$$ \frac{1}{B_1}< t < \frac{1}{D} $$.
For further simplification, we consider the most ``dangerous" situation $\Omega_a=B_b$.
Under this condition, spin $b$ is resonantly driven, so that the expected feedback is the
strongest.  Setting $\Omega_a=B_b$ in (\ref{wide})  we find that the first and the third terms 
vanish.  In the two remaining terms it is sufficient to set $\cos \Omega_bt=1$, since $\Omega_b$
is of the order of $D$.  Also, with $\Omega_a=B_b$, we have $\Omega_b=B_2$.  After that,
(\ref{wide}) simplifies to
\begin{equation}
\label{wide1}
S_z^a(t)S_z^b(t)=  \frac{B_1^2 }{\Omega_a^2}\cos\Omega_a t
 +\frac{\left(\omega - B_a\right)^2}{\Omega_a^2},
\end{equation}
which is nothing but simply $S_z^a(t)$. Still, the behavior of $S_x^b(t)$ emerging upon substitution
of (\ref{wide1}) into (\ref{reduced2}) is nontrivial due to the beating of $\cos \Omega_at$ and $\sin B_b(t-t')$. This beating  generates a contribution to $S_x^b$ equal to $$\frac{B_1^2}{2\Omega_a^2} Dt\sin \Omega_at.$$
We see that this contribution exceeds the free-precession contribution and at $t\sim 1/D$ becomes 
of the order of $1$, which could be expected under the resonant condition $\Omega_a=B_b$.
Our main point is that, even under this condition, the feedback of $S_x^b$ on the first-generation Rabi
oscillations remains small. Indeed, performing the integration $\int\limits_{t'}^t dt''$ in the argument of
cosine in (\ref{reduced1})   generates the correction to this argument equal to
$$\frac{B_1^2D^2}{2\Omega_a^3}\left(t\cos\Omega_at-t'\cos\Omega_at'\right).$$
While this correction grows with $t'$ it does not exceed $1$ as long as $t'$ is smaller than $1/D$.
Thus. we conclude that when the drive exceeds the interaction magnitude, the feedback effect is
negligible.

%
\section{Discussion}

{\em i}. In a driven system of non-interacting spins
in a random magnetic field (random $B_a$), only resonant  spins respond to the drive. With Rabi frequencies depending
on $B_a$, Rabi oscillations of different spins average out,
so that the average $\langle S_z^a(t)\rangle$ approaches 
a constant. We note that this approach is accompanied by
slow-decaying oscillations. Indeed, for a given spin, the
oscillating part $S_z^a(t)$  has the form
\begin{equation}
\label{part}
S_z^a(t)-\overline{S_z^a(t)}=\frac{B_1^2 \cos \left[B_1^2+(\omega-B_a)^2 \right]^{1/2}t}{B_1^2+(\omega-B_a)^2}.
\end{equation}
Assuming the homogeneous distribution of $B_a$ in the 
interval $0<B_a<\Delta$, the disorder-average of (\ref{part}) has the form 
\begin{equation}
\label{disorder}
\langle S_z^a(t)-\overline{S_z^a(t)}\rangle=\frac{1}{\Delta}\int\limits_0^{\Delta} dB_a \frac{B_1^2 \cos \left[B_1^2+(\omega-B_a)^2 \right]^{1/2}t}{B_1^2+(\omega-B_a)^2}.
\end{equation} 

At long times, $B_1t\gg 1$, only the resonant  spins contribute to the integral. This allows to expand the 
argument of cosine as
\begin{equation}
\label{disorder2}
\left[B_1^2+(\omega-B_a)^2 \right]^{1/2}t\approx B_1 t+\frac{(\omega-B_a)^2}{2B_1}t.
\end{equation}
We see that the relevant domain of $\left(\omega-B_a\right)$ is $\sim \left(\frac{B_1}{t}\right)^{1/2}\ll B_1$.
This allows us to set $B_a=\omega$ in the denominator of (\ref{disorder}) and to extend the integration domain
over $(\omega-B_a)$
to $(-\infty, \infty)$. Performing the Gaussian integration, we get
\begin{equation}
\label{disorder3}
\langle S_z^a(t)-\overline{S_z^a(t)}\rangle\Big|_{B_1t\gg 1}= \left(\frac{\pi B_1}{\Delta^2t}\right)^{1/2}\cos\left(B_1t+\frac{\pi}{4}\right).
\end{equation}
In Fig. \ref{f2} the result of numerical calculation of $\langle S_z^a(t)\rangle$ 
for a certain parameter set is shown. Numerics confirms the presence of slow-decaying
periodic oscillations in average $\langle S_z^a\rangle$. 
The amplitude of these oscillations of average $S_z(t)$ coming from the sparse resonant spins
should be compared to the $S_z^b$ coming from the typical second-generation Rabi oscillations 
(\ref{Result1}). The reasonable choice of $\Delta$ is $\omega$. Characteristic $t$ in (\ref{Result1}) is $\sim 1/D_0$. Then $S_z^b$ is $\sim ga^3D_0$, while the oscillating part of
$\langle S_z^a(t)\rangle$ can be presented as $\sim \frac{B_1}{\omega}\left(\frac{D_0}{B_1}\right)^{1/2}$, which is the product of two small parameters. On the other hand, the fact that we considered the interaction of spin $a$ with
only one spin $b$, requires that the product $ga^3D_0$ is also small.



{\em ii.} Spin $b$ can induce even slower Rabi nutations in spin $c$, see Fig. \ref{frabi}. The corresponding Rabi frequency of these third-generation oscillations will be   
\begin{equation}
\Omega_c=\left[B_3^2 + (\Omega_b-B_c)^2 \right]^{1/2},
\end{equation}
where $B_3$ is given by
\begin{equation}
 B_3=\frac{D_{bc} B_2^2}{2 \Omega_b^2}=\frac{D_{bc}D_{ab}^2 B_1^4}{8\Omega_a^4\Omega_b^2}.
\end{equation}
\begin{equation}
 B_3=\frac{D_{bc}(D_{ab})^2 B_1^4}{8(\omega-B_a)^4(\omega-B_a-B_b)^2}
\end{equation}
If we are away from resonance at each step, then drive 
amplitude at $n$-th step will be
\begin{equation}
 B_n=\frac{D_{n,n-1}B_{n-1}^2}{2\Big(\omega-{\sum\limits_{\s i=\left\{a,b,c...n-1~\text{terms}\right\}}} B_i\Big)^2}.
\end{equation}
By contrast, if we are at resonance in each step, then the drive 
amplitude for the $n$-th step will depend only on the dipole-dipole interaction between $n$-th spin and the $n-1$-th spin. For example, the drive amplitude for $n=3$ is $D_{bc}/2$, and the drive amplitude for $n=2$ is $D_{ab}/2$, see Fig. \ref{frabi}. 

Suppose that spin $a$ is not in resonance with the drive, $(\omega-B_a)> B_1$. Then the amplitude of the first-generation Rabi oscillations is small. Still, spin $b$ can oscillate with big amplitude, $\sim 1$,
provided that $B_b$ is equal to $\Omega_a\approx (\omega-B_a)$. At the same time, the frequency of the oscillations
of spin $b$ will be approximately $\frac{D_{ab} B_1^2}{2 (\omega-B_a)^2}$, which is much smaller than $D_{ab}$.

{\em iii.}  It follows from Eq. (\ref{Result1}) that, while the contribution of the second-generation Rabi oscillations to the absorption
is a slow function of time, the magnitude of this slow component
contains a small parameter $ga^3D_0$. Thus, as the dipole-dipole interactions increases, the amplitude of the slow component grows
linearly with $D_0$. On the other hand, the characteristic time before they saturate drops
as $1/D_0$.

{\em iv.} 
For slow second-generation Rabi oscillations to develop
the drive amplitude should be bigger than the interaction strength. On the other hand, the drive is
assumed to be weak,  $B_1\ll \omega$,  which, in classification of
Ref. \onlinecite{Chalker}, corresponds to the linear absorption regime. Such a weak drive cannot affect the 
overall many-body localized 
regime.\cite{PRL2,PRL1,Science1,Science2,Science3}

{\em v.}
 In spirit, the relay-race mechanism considered in the present paper  
 bares some similarity to the mechanism of delocalization of eigenmodes of
 dipole-dipole coupled oscillators or of the ensemble of two-level systems. \cite{Levitov1,Levitov2,Levitov3,Kagan} In Refs. \onlinecite{Levitov1,Levitov2,Levitov3,Kagan}  two undriven  
 oscillators or two spins get hybridized when the corresponding frequencies
 match each other within the interaction magnitude. 
This hybridization can be mediated by $z$-$z$ component of the interaction.
In our notations, the frequencies of two hybridized oscillators can be expressed as
\begin{equation}
\label{levitovfrequency}
\omega^2=\frac{B_a^2+B_b^2}{2}\pm \Big[\frac{\left(B_a^2-B_b^2\right)}{4}+D_{ab}^2 \Big]^{1/2}.
\end{equation}
 We see that, even at resonance $B_a=B_b$,
  hybridization does not result in a slow motion.
  By contrast, in our situation, the resonance is
  dictated by drive and hybridization takes place
  when $\Omega_a$ is close to $B_b$.
In other words, the motion of $a$ {\em in the ``rotated" frame} is in resonance with $b$
in the lab frame.

{\em vi.} We introduced the relay-race mechanism using the language of spins driven
by ac magnetic field. In Refs. \onlinecite{Chalker}, \onlinecite{Demler1}, and \onlinecite{Bhatt} the ac absorption of electric field by localized electrons has
been studied. The main difference between the two scenarios is that we considered the
fields $B_a$ to be random, but parallel to $z$. In the case of the ac electric field,
$\bm{{\cal E}}\cos\omega t$, the Hamiltonian describing the drive has the form
$\bm{P\cdot{\cal E}}\cos \omega t$, where $\bm{P}$ is the dipole matrix element between
the ground and excited states. In spin language, randomness of the directions of $\bm{P}$
translates into the randomness of the directions of $B_a$. In a general
case when ${\bf B}_a={\bf n}_a B_a$, ${\bf B}_1={\bf n}_1B_1$, where ${\bf n}_a$ and ${\bf }n_ 1$ are the unit vectors, the drive amplitude in the above expressions should be modified as
\begin{equation}
B_1^2\rightarrow  B_1^2
\left({\bf n}_1\times{\bf n}_a\right)^2. 
\end{equation}
It is important to note that when the directions of the fields ${\bf B}_a$ are random, we
do not need the non-diagonal component of dipole-dipole interaction to induce the
second-generation Rabi oscillations.

{\em vii.} There is a similarity between the relay-race mechanism considered above and the Rabi-vibronic resonance studied
in Ref. \onlinecite{Glenn}. In the latter case,= the Rabi oscillations
are resonantly coupled to a vibronic mode rather than to the neighboring spin. 

\vspace{4mm}

\centerline{\bf Acknowledgements}
The work was supported by the Department of
Energy, Office of Basic Energy Sciences, Grant No. DE-
FG02-06ER46313.

\end{document}